\documentclass{basi}
\usepackage{epsfig}
\begin{document}
\title[Mass limit on Nemesis]{Mass limit on Nemesis}
\author[Varun Bhalerao and M.N. Vahia]%
{Varun Bhalerao{$^1$\thanks{email:varunb@iitb.ac.in}} and M.N. Vahia{$^2$\thanks{email: vahia@tifr.res.in}}\\
$^1$ Indian Institute of Technology, Bombay, Powai, Mumbai 400 076, \\
$^2$ Tata Institute of Fundamental Research, Homi Bhabha Road, Colaba, Mumbai 400 005}
\date{Received; Accepted 2005 February 10} 
\maketitle
\label{firstpage}
\begin{abstract}
We assume that if the sun has a companion, it has a period of 27 Myr 
corresponding to the periodicity seen in cometary impacts on earth. Based on this 
assumption, it is seen that the inner Lagrengian point of the interaction 
between the Sun and its companion is in the Oort cloud. From this we calculate 
the mass -- distance relation for the companion. We then compute the expected 
apparent magnitude (visible and J band) for the companion using the models of 
Burrows (1993). We then compare this with the catalogue completeness of optical 
and infrared catalogues to show that the sun cannot have a companion of mass 
greater than 44 M$_{jup}$ (0.042 M$_{sun}$)
\end{abstract}


\section{Introduction}
\label{sec:intro}
About half the stars in the galaxy are in binaries (Harwitt, 1998). The presence 
of a possible companion to the Sun has often been speculated about and has 
been named as `Nemesis' in the literature (see for example Bailey, 1984). These 
speculations have been based on the observations of approximate periodicity in 
mass extinction on earth, possibly due increasing frequency of cometary impact. 
However, the current observations have not found any possible solar companion 
and only indirect limits have been placed on the mass of Nemesis based on 
interpretation of geological records. 
Raup and Sepkoski (1984) show that there is a periodicity in mass extinctions on the earth, and they attribute this to cometary impacts. They assume 
that a binary to the sun perturbs the Oort cloud which increases the number of 
near earth comets and therefore increases the probability of a cometary impact 
on the Earth. 
The only search for Nemesis so far was conducted at Berkeley, but was 
stopped soon (Muller, 2004). Carlson et. al. (1994) have hypothesized that 
Nemesis may be a red dwarf. 

The most extensive work on this problem has been done by Matese, 
Whitman and Whitmire (1998) proposed a mass of about 3 M$_{jup}$ for this object. 
They have calculated the trajectories of 82 newly discovered comets and find 
that 25$\%$ of them come from a well defined location in the sky. They therefore 
assume that a perturber sitting in the Oort's cloud at 25,000 AU and fed by 
Galactic tide can adequately explain the various properties of the observed 
comets. However, in order for the object to be effective, the Oort's cloud has 
to be perturbed by the galactic tide. 

In the present paper we try to generalize the constraints of the nature 
of the binary companion of the Sun. We assume that the Oort's cloud is stable 
and comets are regularly perturbed into the inner solar system due to external 
perturbations. We assume that this happens because the Inner Lagrangian 
Point (L1) of the Sun -- Nemesis system must be at the Oort cloud. Using this, we 
attempt to estimate the upper limit on the mass -- distance relation of Nemesis. 
We then compare this data with the observational catalogues and show that 
the mass limits on the solar companion can be quite severe. 

\section{Assumptions}
 
It has been suggested there is a 27 million year periodicity in the arrival of long period comets to the Earth (Raup and Sepkoski, 1984) based on extinction rate of species and other geological evidence of cometary impact. If this is true then the Oort's cloud objects would  be perturbed as they moved in the region close to the inner  Lagrangian point between the Sun and Nemesis. We therefore assume that the inner Lagrangian point L1 of the Sun - Nemesis system passes through the Oort cloud at different distances (see e.g. Muller, 2002). Based on this, we derive the mass  distance relation for the companion. We then attempt to calculate the apparent luminosity of the companion. We assume that the object is as old as the sun and ignore heavier stars since their life times are much shorter (Bowers and Deeming, 1984). It is also seen from Figure 1 that the apparent magnitude of heavier objects under given constraints would be lower than +4 which would make them visible even to the unaided eye. For the smaller masses, we use the models from Burrows et al. (1993) and Burrows et al. (1997). We assume the evolution of the small mass bodies along the path suggested therein, and neglect the case where the object could be a low mass star at the lower end of the main sequence. Burrows (1997) (see figure 7 in the Burrows, 1997) has shown that for objects of mass less than 0.2 M$_{sun}$ till 0.0003 M$_{sun}$, the fall in the luminosity is extremely severe after 10$^{9}$ years and the magnitude for the maximum to minimum mass in this range is of the order of 8 orders of magnitude. However the absolute magnitude for 0.0003 M$_{sun}$ object after 10$^{9.5}$ years is -10 which defines the lower limit of the sensitivity of the present work. From these we calculate absolute visible and J band magnitude of the companion assuming that it radiates as a blackbody. We ignore the case of neutron star or black hole companion since these objects illuminated by accretion will have very high intrinsic luminosity.

\section{Calculations}
We calculate the distance to the companion by iteratively solving the equation for the Lagrangian point between Sun and Nemesis as follows. At the inner Lagrangian point we write the force balance equation

\begin{equation}
\label{eq1}
\frac{G M_s}{d^2} = {\frac{GM_n}{(r-d)^2}} + (d-r_1)G\frac{M_n+M_s}{r^3}  
\end{equation}

Here $M_s, M_n$ are the masses of Sun and Nemesis, $r$ is the separation between Sun and Nemesis, $d$ is the distance of L1 from the Sun $r_1$ is the distance of center of mass of the Sun $-$ Nemesis system from the Sun (given by $\frac{rM_s}{M_s+M_n}$. This can be simplified as

\begin{equation}
\label{eq2}-(m+1)x^5+(2m+3)x^4-(m+3)x^3+mx^2-2mx+m=0 
\end{equation} In the equation \ref{eq2}, $m = \frac{M_s}{M_n}$ and $x = \frac{d}{r}$.On the basis of this equation we derive the mass distance relation for a solar companion (figure \ref{fig2}).

We adopt the radius and temperature of these objects from models from Burrows et al. (1993) and Burrows et al. (1997). We note here that the evolutionary models used in Burrows et al. (1997) are based only on initial mass and hence do not involve any classification into "stars", "brown dwarfs" and "planets". We have derived the apparent magnitudes from this data using our mass$-$distance relations. The apparent J band and visible magnitudes are given in table \ref{tab1} for objects of different masses. We have taken several sample masses from about 0.0005 M$_{sun}$ to about 0.24 M$_{sun}$.

\begin{table}
\centering
\small\caption{Mass distance relation for Nemesis for period 27 million years.}
\label{tab1}
\begin{tabular}{cccc} \\
\hline\hline 
M$_{comp}$/M$_{sun}$&Pair separation&Apparent&Apparent\\
&(AU)& V magnitude &J magnitude\\\hline
0.004&90,114&122.9&68.7\\
0.007&90,199&95.2&54.1\\
0.010&90,297&76.3&44.2\\
0.015&90,454&59.5&35.3\\
0.020&90,594&50.4&30.5\\
0.024&90,707&45.5&27.9\\
0.027&90,791&42.2&26.2\\
0.028&90,847&40.0&25.0\\
0.030&90,886&38.2&24.1\\
0.035&91,032&34.2&22.0\\
0.040&91,178&30.3&20.0\\
0.045&91,323&27.2&18.3\\
0.050&91,468&23.8&16.5\\
0.055&91,612&22.3&15.8\\
0.060&91,756&20.2&14.7\\
0.065&91,899&18.2&13.6\\
0.071&92,070&15.2&12.0\\
0.076&92,213&11.4&10.0\\
0.081&92,355&7.3&7.9\\
0.086&92,496&4.9&6.6\\
0.091&92,637&4.1&6.1\\
0.095&92,778&3.7&5.8\\
0.099&92,890&3.4&5.7\\
0.149&94,270&1.8&4.5\\
0.199&95,611&1.1&3.9\\
0.236&96,579&0.7&3.6\\
\hline\end{tabular}\smallskip\end{table}\

\begin{figure*}
   \centering
   \includegraphics[scale=0.5]{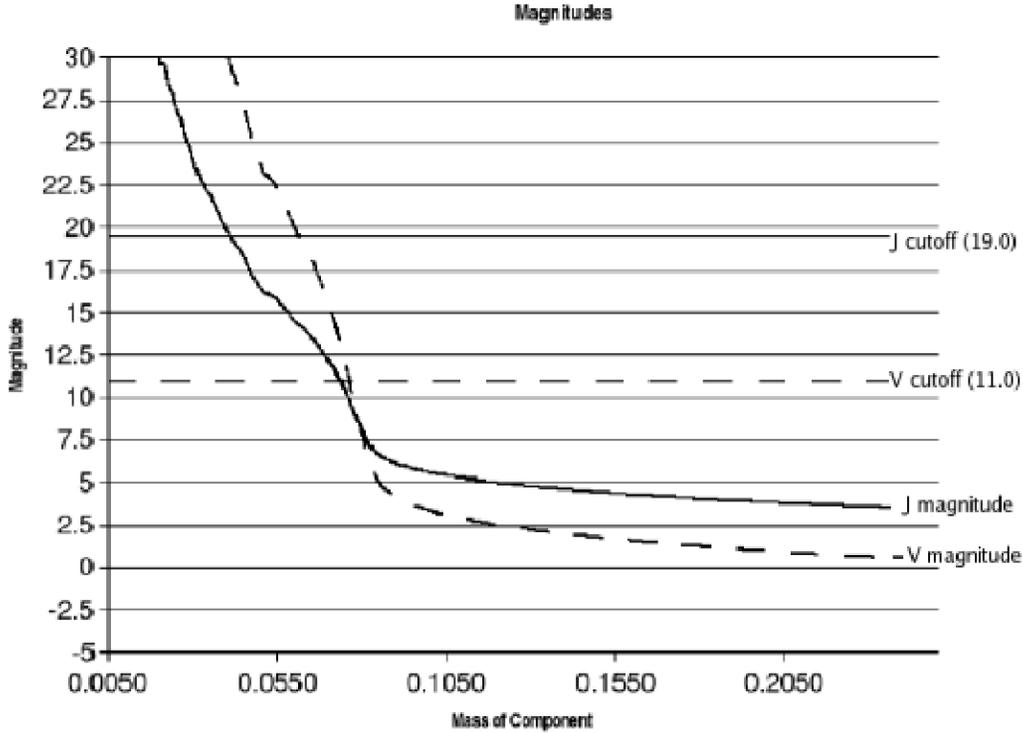}
   \caption{Magnitude versus Mass graph for Nemesis for a assumed period of 27 MYr. The horizontal lines indicate the IR and optical catalogue limits}
   \label{fig1}
\end{figure*}

The first column in table 1 is the mass of Nemesis in terms of Solar mass, the second column is the distance of assuming a period of 27 Myr (see figure \ref{fig1}), and columns 3 and 4 give the apparent visual (400-700 nm) and infrared (J band 1.24 $\pm$ 0.1 micron) magnitudes (ESO, 2005).

\eject

We also calculate the change in apparent magnitudes (and hence the shift in cutoff mass) as 
result of change in the period. Calculations show that by increasing the period by 1 Myr increases the apparent magnitude (visible as well as J band) by about 0.05, while decreasing 
the period by 1 Myr decreases the apparent magnitude (visible as well as J band) by about 0.05.

\section{Discussion}
\begin{figure}[h]
   \centering
\hspace{-1cm}   
   \includegraphics[scale=0.5]{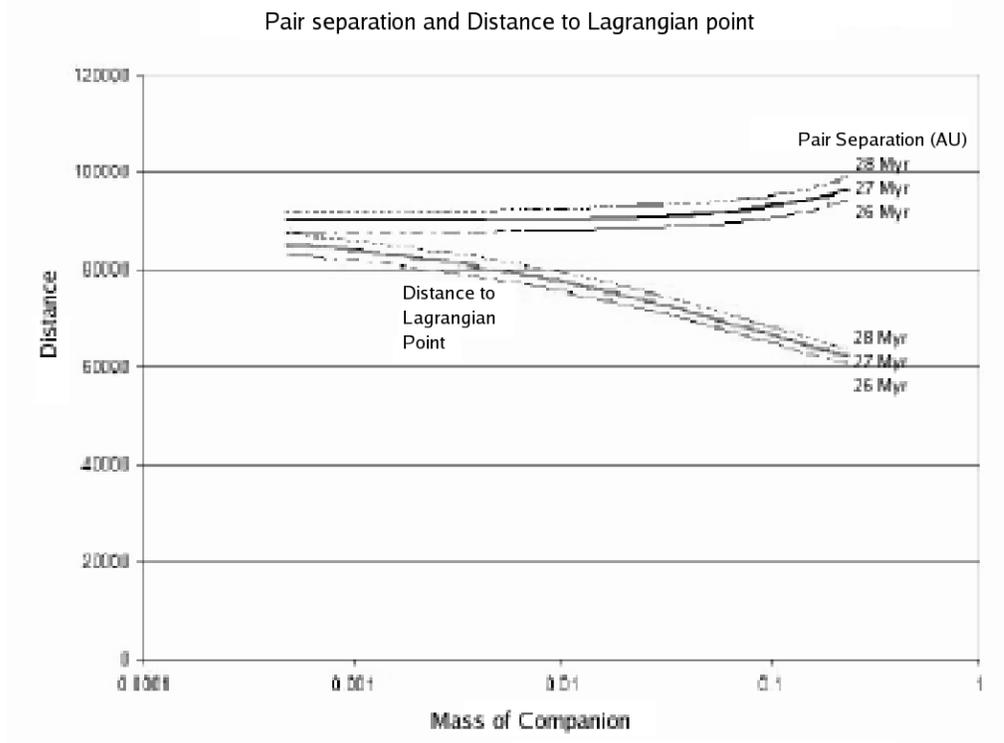} 
   \caption{Distance from the Sun to L1 and to Nemesis as a function of mass of Nemesis,for values of the orbital period of nemesis as 26 Myr, 27 Myr and 28 Myr.}
   \label{fig2}
\end{figure}

We attempt to estimate the limit on the mass of Nemesis based on the above 
calculations. Figure \ref{fig2} shows the distance from the Sun to L1 and to Nemesis 
as a function of mass of Nemesis. The curves are drawn for values of the orbital 
period of nemesis as 26 Myr, 27 Myr and 28 Myr.

In figure \ref{fig1} we plot the apparent magnitudes as a function of the mass of nemesis, 
for a period of 27 Myr. We note that calculations have shown that magnitude varies 
only by a small amount for even a 1 Myr change in period. The horizontal lines 
indicate the catalogue completeness as defined below. 
The Tycho 2 star catalouge (Hog et al., 2000) is complete till m$_v$ =11.0 
(see also Vizier Catalogue service). The Guide Star Catalouge GSC 2.2 taken 
from STScI(2001) is complete to J= 19.5. 
From figure 2 and the calculations we conclude that the sun cannot have 
an unobserved companion with mass $>$ 44 M$_{jup}$. 

We estimate the error in this limit as follows. Since the periodicity in 
the geological records is 27 million years, the sum of the perihelion and aphelion 
distance must be 180,000 AU from Kepler's laws approximated for a low mass 
companion. Hence, if the object is in a highly elliptical orbit, the farthest the 
object can be, is 180,000 AU from the Sun. If the object is presently at its 
aphelion distance, then the apparent brightness will be a factor of 4 less than 
the value calculated here, effectively increasing the apparent magnitude by 1.5. 
This shifts the cutoff to about 0.045 M$_{sun}$ (47 M$_{jup}$).

Lopatnikov et al. (1991) estimate the Oort cloud mass to be about 300 earth masses 
(about 0.95 M$_{jup}$), while in more recent work of Weissman (1996) estimates the Oort 
cloud mass to be about 38 earth masses (0.12 M$_{jup}$). We note that this mass will 
be distributed through the entire Oort cloud. So, its influence on the orbit of 
a companion of $>$ 40 M$_{jup}$ (which is our range) can be neglected safely.
\section{Conclusion}
We conclude that if the Sun$-$ Nemesis system has a period of 27 Myr, the sun cannot have a companion $>$ 44 $M_{jup}$ (0.042 M$_{sun}$)

\section*{Acknowledgements}
One of us (VB) wishes to thank Chanitanya Ghone and Surhud More for their 
assistance and support. 

The Guide Star CatalogueII is a joint project of the Space Telescope 
Science Institute and the Osservatorio Astronomico di Torino. Space Telescope 
Science Institute is operated by the Association of Universities for Research 
in Astronomy, for the National Aeronautics and Space Administration under 
contract NAS526555. The participation of the Osservatorio Astronomico di 
Torino is supported by the Italian Council for Research in Astronomy. Additional support is provided by European Southern Observatory, Space Telescope 
European Coordinating Facility, the International GEMINI project and the 
European Space Agency Astrophysics Division. 

\label{lastpage}
\end{document}